\documentclass{aipproc}

\layoutstyle{8x11single}
\SetInternalRegister\hbadness{8000} % pseudo latin isn't breaking very well :-)

\newcommand\doingARLO[2][]{%
  \ifx\mmref\undefined #1\else #2\fi
}

\begin{document}

\title 
      {Neutrino induced weak pion production off the
 nucleon and coherent pion
production in nuclei at low energies}

\classification{25.30.Pt, 12.15.-y, 13.15.+g}
\keywords{Neutrino reactions, coherent pion production}

\author{J.E. Amaro}{
  address={Departamento de F\'{\i}sica At\'omica, Molecular y Nuclear,
  Universidad de Granada, E-18071 Granada, Spain}
%  email={frank.mittelbach@latex-project.org},
%  thanks={This work was commissioned by the AIP}
}

\iftrue
\author{E. Hern\'andez}{
  address={Departamento de F\'{\i}sica Fundamental e IUFFyM, Facultad de
  Ciencias, E-37008 Salamanca, Spain},
%  email={arno@mittelbach-online.de},
}

\author{J. Nieves}{
  address={Instituto de F\'{\i}sica Corpuscular (IFIC), Centro Mixto
  CSIC-Universidad de Valencia, Institutos de Investigaci\'on de Paterna,
  Aptd. 22085, E-46071 Valencia, Spain},
%  email={arno@mittelbach-online.de},
}
\author{M. Valverde}{
  address={Research Center for Nuclear Physics (RCNP), Osaka University,
Ibaraki 567-0047, Japan},
%  email={arno@mittelbach-online.de},
}
\author{M.J. Vicente-Vacas}{
  address={Departamento de F\'{\i}sica Te\'orica and IFIC, Centro Mixto
  CSIC-Universidad de Valencia, Institutos de Investigaci\'on de Paterna,
  Aptd. 22085, E-46071 Valencia, Spain},
%  email={arno@mittelbach-online.de},
}
\fi

% \copyrightholder{Acoustical Scociety of America}
\copyrightyear  {2001}

\begin{abstract}
We present a microscopic model   for neutrino induced
one-pion production off the nucleon and its implementation for the purpose of calculating coherent pion 
production in  nuclei.
  We further criticize the use of the Rein--Sehgal model for coherent pion production
by low energy neutrinos. In particular, we show how the approximations in that model
give rise to a much flatter differential cross section in the $\eta=E_\pi(1-\cos\theta_\pi)$
variable.  We discuss the limitations intrinsic to any approach based on
the partial conservation of the axial current hypothesis and the inability of such models
 to properly determine the angular
distribution of the outgoing pion with respect to the direction of the
incoming neutrino. We show the effects of those limitation for the case of the $\frac{d\sigma}{d\eta}$ 
differential cross section. 
%\thanks{A second footnote in the abstract}
\end{abstract}

\date{\today}

\maketitle
%
%
%      Introduction
%
%
\section{Introduction}
Neutrino induced one-pion production off nucleons and nuclei in the
intermediate energy region is a source of relevant data on hadronic
structure. Pions are mainly produced through resonance excitation and
these reactions can be used to extract information on
nucleon-to-resonance axial transition form factors. Besides, a proper
understanding of these processes is very important in the analysis of
neutrino oscillation experiments since pion production could be a source of background
in those experiments \cite{AguilarArevalo:2007it,Hiraide:2006zq}. 

In reactions on nuclei, pions can be produced incoherently or
coherently. In the latter case the nucleus remains in its ground
state. Coherent reactions are controlled by the nucleus form factor
and are more forward peaked than incoherent ones.
Experimental analyses of the coherent reaction rely on the Rein--Sehgal (RS) model~\cite{Rein:1982pf}
which assumes that coherent pion production is dominated by
the divergence of the axial
current
and can thus be related to the pion-nucleus coherent scattering through the
partial conservation of the axial current (PCAC) hypothesis. For instance for coherent $\pi^0$ production the
RS model approximates the coherent cross section for both  neutrino and
antineutrino induced processes by
\begin{equation}
\frac{d\sigma}{dx\,dy\, dt}
= \frac{G^2 M
E}{\pi^2}f_\pi^2 (1-y)\frac{1}{(1-q^2/1\ {\rm GeV}^2)^2} 
\left (|F_{\cal A}(t)|^2 F_{\rm abs}
\frac{d\sigma(\pi^0 N \to \pi^0  N)}{dt}\Big|_{E_\pi=q^0, t=0}\right )\,,
\label{eq:rsNucleon}
\end{equation}
with $G$  the Fermi constant, $M$ the nucleon mass, $E$ the incident neutrino energy and
$f_\pi$ the pion decay constant. Besides $q$ is the lepton four-momentum transfer, 
 $x=-q^2/2Mq^0$, 
$y=q^0/E$, and $t=(q-k_\pi)^2$ with $k_\pi$ the pion four-momentum. $t$ equals $-(\vec{q}-\vec{k}_\pi)^2$ 
if, as usually assumed, the  nucleus
 recoil is neglected ($q^0=k_\pi^0\equiv E_\pi$). $F_{\cal A}(t)$ is the  the nuclear form
factor  calculated as $F_{\cal A}(t)=\int d^3\vec{r}\ e^{{\rm
i}\left(\vec{q}-\vec{k}_\pi\right)\cdot\vec{r}}
\left\{\rho_p(\vec{r}\,)+\rho_n(\vec{r}\,)\right \}$ with $\rho_{p(n)}$ the nuclear
proton (neutron) density, normalized to the number of protons
(neutrons). $F_{\rm abs}$ is a $t-$independent attenuation factor that takes into account the
distortion of the final pion. Finally 
$\frac{d\sigma(\pi^0 N \to \pi^0  N)}{d|t|}\Big|_{q^0=E_\pi, t=0}$ is the differential
pion-nucleon cross section evaluated at $t=0$.
The $t=0$ approximation in the pion-nucleon cross section is not needed and it can  be justified
only if the nuclear form factor is sufficiently forward peaked. 
The larger the pion energy and the heavier the nucleus, the better this approximation becomes. 
In the original paper~\cite{Rein:1982pf}, the model  was
applied to medium size nucleus, aluminum, and neutrino energies
above 2 GeV, for which the relevant pion energies are quite
high. However, as we  pointed out in
~\cite{Amaro:2008hd}, for neutrino energies below 1 GeV and lighter
nuclei, like carbon or oxygen, the nuclear form factor is not enough
forward peaked to render the finite $t-$dependence of the
pion-nucleon cross section negligible. The distortion factor $F_{\rm
abs}$ is also an oversimplification since in any realistic scattering model
this factor should depend on $t$. The recent work of Berger and Sehgal~\cite{Berger:2008xs} already corrects some of
the problems in the RS model.

In Ref.~\cite{Hernandez:2009vm} we discuss in detail the approximations inherent to any
PCAC based model and, in particular, the approximations in the RS model. As we show
below, the neglect 
of non-PCAC terms and the implicit assumption (only correct for $q^2=0$) by  PCAC based models that
%\begin{equation}
$\frac{d\sigma}{dx\,dy\,dt\,d\phi_{k_\pi q}} 
= \frac{1}{2\pi} \frac{d\sigma}{dx\,dy\,dt},
$
%\label{eq:ang-approx}
%\end{equation}
where $\phi_{k_\pi q}$ is the azimuthal pion angle in a plane perpendicular to 
$\vec q$,
affects the determination of the differential cross sections with respect to 
 the angle made by the pion
and the incident neutrino, 
leading to flatter distributions in that variable. 
 In the particular case of the RS model the $t=0$ approximation
enhances this unwanted effect. 
This could have implications for the recent determination by the MiniBooNE
Collaboration of the neutral current (NC) coherent $\pi^0$ production rate~\cite{AguilarArevalo:2008xs}.

Our approach to the problem, as others in the literature, does not rely on PCAC. We use a microscopic model in which
coherent pions are mainly produced by virtual $\Delta$-hole excitations in the nucleus, 
as well as  additional 
processes that are required by chiral symmetry. The model includes the
modifications of the  $\Delta$ propagator in the nucleus by means of a sophisticated evaluation 
of the $\Delta$ self-energy in the medium, and it treats the final pion distortion in a realistic way 
by solving the Klein-Gordon (KG) equation for a pion-nucleus optical potential.
% In the following sections we
%briefly explain the microscopic model at the nucleon level and its implementation for coherent pion production
%in nuclei.
%
%          Elementary model
%
%
\section{Microscopic model at the nucleon level}
\begin{figure}[tbh]
\centerline{\includegraphics[height=7cm]{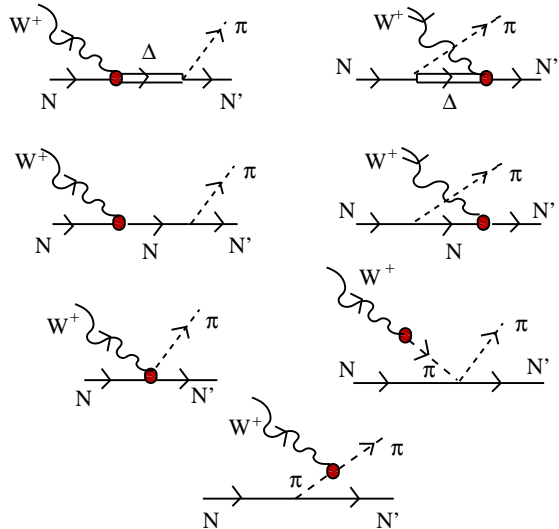}}
\caption{\footnotesize Model of Ref.~\cite{Hernandez:2007qq} for the $W^+N\to
  N^\prime\pi$ reaction.  The circle in the diagrams stands for the
  weak transition vertex. }\label{fig:diagramas}
\end{figure}

The model that we use to describe the elementary production  process at the nucleon level
is the one derived in Ref.~\cite{Hernandez:2007qq}. In  
Fig.~\ref{fig:diagramas} we give the different contributions for the charged current (CC)
$W^+N\to N'\pi$ case
(details for the antineutrino induced process and the NC case can be found in~\cite{Hernandez:2007qq}). In addition to the dominant $\Delta$ pole
($\Delta P$) mechanism (weak excitation of the $\Delta(1232)$
resonance and its subsequent decay into $N\pi$), the model also includes
the crossed $\Delta$ pole term ($C\Delta P$) and other background terms that
are required by chiral symmetry:  Direct and crossed
 nucleon (second row) pole terms
( $NP$, $CNP$), contact ($CT$) and pion pole
($PP$) contribution (third row) and finally the pion-in-flight ($PF$)
term. 
In
Ref.~\cite{Hernandez:2007qq}, we found that background terms produced
significant effects in all channels.  The least known
ingredients of the model are the axial nucleon-to-$\Delta$ transition
form factors, of which $C_5^A$ gives the largest contribution (See Eq.(1) of Ref.~\cite{Schreiner:1973ka}
 for a form factor  decomposition of the $N\Delta$ weak current). Besides, within
the Adler model that we use~\cite{adler}, $C_5^A$ determines all other axial
form factors. This
strongly suggested the readjustment of that form factor to the experimental
data, which we did by fitting the flux-averaged $\nu_\mu p\to \mu^- p
\pi^+$ ANL~\cite{Barish:1978pj,Radecky:1981fn} $q^2$-differential cross section for pion-nucleon invariant
masses $W < 1.4$ GeV, for which the model should be appropriate. 
Assuming the shape~\cite{Pa04} 
\begin{equation}
C_5^A(q^2) = \frac{C_5^A(0)}{(1-q^2/M^2_{A\Delta})^2}\times
\frac{1}{1-\frac{q^2}{3M_{A\Delta}^2}}\label{eq:ca5old}
\end{equation}
we found $C_5^A(0) = 0.867 \pm 0.075$ and $M_{A\Delta}=0.985\pm 0.082\,{\rm
  GeV}$.
Our full
model, thus obtained, lead to an overall better description of the
data for different CC and NC, neutrino- and antineutrino-induced,
one-pion production reactions off the nucleon~\cite{Hernandez:2007qq}. 
The
$C_5^A(0)$ value is significantly smaller  than the
traditionally used value of about $1.2$ deduced from the off-diagonal
Goldberger--Treiman relation. This reduction of 
$C_5^A(0)$  is consistent with recent results in lattice QCD
\cite{Alexandrou:2006mc}, quark model \cite{BarquillaCano:2007yk} and
phenomenological studies \cite{Graczyk:2007bc}.
%
%
%
%  Coherent production in nuclei
%
%
\section{Coherent production in nuclei: model and results}

Now the process consists of a weak pion  production followed by the strong
distortion of the pion in its way out of the nucleus. In the coherent
production the nucleus is left in its ground state and the hadronic amplitude
 is written as a
 sum over the amplitudes for each nucleon. Neglecting for the moment
 nonlocalities  and
pion distortion one can write for the hadronic amplitude 
\begin{eqnarray}
 {\cal A}^\mu_{\pi}(q,k_\pi) =  \int d^3\vec{r}\ e^{{\rm
        i}\left(\vec{q}-\vec{k}_\pi\right)\cdot\vec{r}}
    \left\{\rho_p(\vec{r}\,) 
\Big[{\cal J}^\mu_{p\,\pi}(q,k_\pi)\Big] 
+ \rho_n(\vec{r}\,) 
\Big[{\cal J}^\mu_{n\,\pi}(q,k_\pi)\Big] \right \}  
\label{eq:Jmunu2}
\end{eqnarray}
where the approximation $q^0=k_\pi^0$ (neglect of nucleus recoil) has been used. One can see from
Eq.(\ref{eq:Jmunu2}) that
the coherent pion production process is sensitive to
the Fourier transform of the nuclear density for momentum
$\vec{q}-\vec{k}_\pi$. ${\cal
  J}^\mu_{N\pi}(q,k_\pi)$ in Eq.(\ref{eq:Jmunu2}), stands for the nucleon spin
averaged $WN\to N \pi$ or $Z^0N \to N \pi$ weak transition amplitude 
\begin{equation}
{\cal  J}^\mu_{N\pi}(q,k_\pi)=\frac{1}{2} \sum_r \bar u_r(\vec{p}^{\,\prime}) \Gamma_{i; N\pi}^\mu
u_r(\vec{p}\,)\,\frac{M}{p^0},
\quad i=\Delta P,\, C\Delta P,\, NP,\, CNP,\, CT,\, PP,\, PF \label{eq:avg}
\end{equation}
where the $u$'s are Dirac spinors for the nucleons, normalized such
that $\bar u u=2M$, and the four-vector matrices $\Gamma_{i;
  N\pi}^\mu$ can be read from the explicit expressions of the pion
production amplitudes given  in  Ref.~\cite{Hernandez:2007qq}.  $\vec{p}$ and
$\vec{p}^{\,\prime}=\vec{p}+\vec{q}-\vec{k}_\pi$ are the initial and
final three momenta of the nucleon. Those momenta are not well defined
and we approximate the four-momentum of the initial nucleon  by
%
%\begin{equation}
$p^\mu = \sqrt{M^2+ \frac14{(\vec{k}_\pi-\vec{q})^2}} ,
  \frac{\vec{k}_\pi-\vec{q}}{2}\,)$.  
%\label{eq:pmu} \, ,
%\end{equation}
Hence we assume that the initial nucleon momentum is
$(\vec{k}_\pi-\vec{q}\,)/2$ and the final one is
$-(\vec{k}_\pi-\vec{q}\,)/2$, with both nucleons being on-shell.  The
momentum transfer is equally shared between the initial and final
nucleon momenta.
Setting $\vec{p}=-\vec{p}^{\,\prime}=(\vec{k}_\pi-\vec{q}\,)/2$ eliminates
some non-local contributions, and it greatly simplifies the sum over all
nucleons, which can be cast in terms of the neutron and proton
densities. Furthermore, the sum over
spins in Eq.~(\ref{eq:avg}) can be also easily performed for
$\vec{p}=-\vec{p}^{\,\prime}$ since
$u_r(\vec{p}^{\,\prime}=-\vec{p}\,)= \gamma^0 u_r(\vec{p}\,)$, so that
\begin{equation}
\frac{1}{2} \sum_r \bar u_r(\vec{p}^{\,\prime}=-\vec{p}\,) \Gamma_{i; N\pi}^\mu
u_r(\vec{p}) = \frac{1}{2} {\rm  Tr}\left((\slash\hspace{-.2cm}{p}+M)\gamma^0\Gamma_{i; N\pi}^\mu \right),
\quad i=\Delta P\,, C\Delta P\,, NP\ldots \label{eq:avg2}
\end{equation}
Given the importance of the $\Delta-$pole contribution and since the
$\Delta$ properties are strongly modified inside the nuclear
medium
we consider some additional nuclear corrections to this contribution to
include the effect of the self-energy of the $\Delta$ in the medium
$\Sigma_\Delta(\rho(\vec{r}\,))$. 
We follow the same approach as
in Ref.~\cite{AlvarezRuso:2007tt}, which is based on the findings of
Refs.~\cite{Oset:1987re,Nieves:1993ev,Nieves:1991ye} . 
Thus in the
$\Delta-$propagator, we make the substitutions $M_\Delta\to M_\Delta +
{\rm Re}\Sigma_\Delta $ and $\Gamma_\Delta/2 \to \Gamma_\Delta^{\rm
  Pauli}/2- {\rm Im} \Sigma_\Delta$ and take
$\Sigma_\Delta(\rho(\vec{r}\,))$ and $\Gamma_\Delta^{\rm Pauli}/2$ as
explained in Sect. II-B of Ref.~\cite{AlvarezRuso:2007tt}. Once these corrections are included
${\cal  J}^\mu_{N\pi}$ depends on $\vec r$ as well and the nuclear form factor
can  no longer be factorized out as in the RS model.

So far the formalism has used the bound wave functions of the nucleons in the nucleus,
which appear via the proton and neutron densities, and has considered
only a plane wave for the pion. Pion distortion effects are important,
specially for $|\vec{k}_\pi| < 0.5$ GeV,
and are considered  by replacing in
Eq.~(\ref{eq:Jmunu2})
\begin{eqnarray}
e^{-{\rm i}\vec{k}_\pi\cdot\vec{r}} \to  \widetilde{\varphi}_{\pi}^{\ast}
    (\vec{r};\vec{k}_\pi) \label{eq:pi-nolocal}  \\
\vec{k}_\pi e^{-{\rm i}\vec{k}_\pi\cdot\vec{r}} \to {\rm i}
\vec{\nabla} \widetilde{\varphi}_{\pi}^{\ast}
    (\vec{r};\vec{k}_\pi) \label{eq:pi-nolocal2}
\end{eqnarray} 
The pion wave function $\widetilde{\varphi}_{\pi}^{\ast}
(\vec{r};\vec{k}_\pi)$  corresponds to an incoming solution
of the KG equation,
\begin{equation}
\left [- \vec{\bigtriangledown}^{2} + m_\pi^{2} + 2 E_\pi V_{\rm opt}
(\vec{r})\right ] 
\widetilde{\varphi}_{\pi}^{\ast}
    (\vec{r};\vec{k}_\pi) =
E_\pi^{2} \widetilde{\varphi}_{\pi}^{\ast}
    (\vec{r};\vec{k}_\pi) \,, \label{eq:optical}
\end{equation}
with $V_{\rm opt} (\vec{r})$ the optical potential which describes the
$\pi$-nucleus interaction.  This potential has been developed
microscopically and it is explained in detail in
Refs.~\cite{Nieves:1993ev,Nieves:1991ye}. 
This theoretical potential reproduces fairly well the data of pionic
atoms (binding energies and strong absorption
widths)~\cite{Nieves:1993ev} and low energy $\pi$--nucleus
scattering~\cite{Nieves:1991ye}. At low pion energies, 
it is an improvement over the one used
in \cite{AlvarezRuso:2007tt,AlvarezRuso:2007it}, that was based on
$\Delta$ dominance of the $\pi N$ interaction. 
The replacement in Eq.~(\ref{eq:pi-nolocal2}), that takes into
account the fact that the pion three momentum is only well defined
asymptotically when the pion-nucleus potential vanishes, induces some
non-localities in the amplitudes. To treat these non-localities we
have adopted the following scheme:
\begin{itemize}
\item In the $\Delta P$, $C\Delta P$, $NP$, $CNP$ terms, we note that
  there exist either a $NN\pi$ or a $N\Delta \pi$ vertex (see Eq.~(51)
  of Ref.~\cite{Hernandez:2007qq}), which induces a factor
  $k_\pi^\alpha$ in the amplitudes. Indeed, for those terms we could
  re-write
\begin{equation}
{\cal J}^\mu_{i;N\pi}(\vec{r};q,k_\pi) = 
(k_\pi)_\alpha \hat{\cal {J}}^{\mu\alpha}_{i;N\pi}(\vec{r};q,k_\pi) , 
\quad i=\Delta P, C\Delta P\,, NP\,, CNP\, . 
\label{eq:prescription}
\end{equation}
We do not consider any non-locality in the tensor $\hat{\cal
{J}}^{\mu\alpha}_{i;N\pi}$, and we use the prescription of
Eqs.~(\ref{eq:pi-nolocal}) and (\ref{eq:pi-nolocal2}) to account for
$\vec{k}_\pi$ in the contraction between $k_\pi^\alpha$ and $\hat{\cal
{J}}^{\mu\alpha}_{i;N\pi}$ in Eq.~(\ref{eq:prescription}). For $k_\pi^0$ we shall use the asymptotic pion
energy. This
approach to treat the non-localities is equivalent to that assumed in
Refs.~\cite{AlvarezRuso:2007tt,AlvarezRuso:2007it}.

\item We do not consider any non-locality for the $CT$ and $PP$ contributions.
\end{itemize}
\begin{figure}
  \resizebox{8.5cm}{!}{\includegraphics{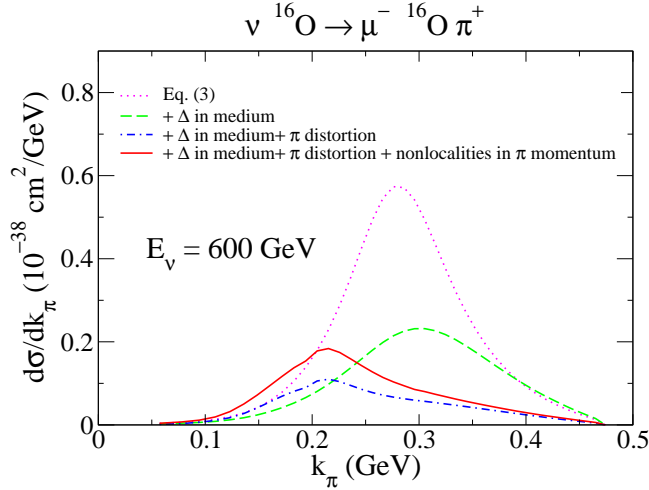}}
\caption{Effects of $\Delta$ in medium, $\pi$ distortion and nonlocalities in
$\pi$ momentum on the $\frac{d\sigma}{dk_\pi}$ differential cross section evaluated for the 
$\nu~^{16}O\to\mu^-  ~^{16}O \pi^+$ reaction at
$E_\nu=0.6\,$GeV.}
\label{fig:medcor}
\end{figure}
The effects of the modifications of the $\Delta$ properties in the medium, pion
distortion and nonlocalities in the pion momentum are shown in
Fig.~\ref{fig:medcor}. As can be seen all corrections are very relevant. Very
recently, it has been claimed that nonlocalities in the nucleon momentum could be
as important, giving rise to a large reduction of the coherent
cross section~\cite{Leitner:2009ph}. The calculation in Ref.~\cite{Leitner:2009ph} has been
done  without $\Delta$ in medium corrections or pion distortion and further
studies
 are needed to clarify this point.

As mentioned before, background terms  gave an important contribution  at the nucleon
level~\cite{Hernandez:2007qq} but they turned out to be irrelevant for coherent production
in symmetric nuclei~\cite{Amaro:2008hd}, and thus the reduction we found in $C_5^A(0)$ amounts to an important
 decrease
in the coherent cross section. In Fig.\ref{fig:cross} we give
cross sections for CC  and NC coherent pion production
on carbon and oxygen. Other observables are discussed in Ref.~\cite{Amaro:2008hd}.
\begin{figure}[htb]
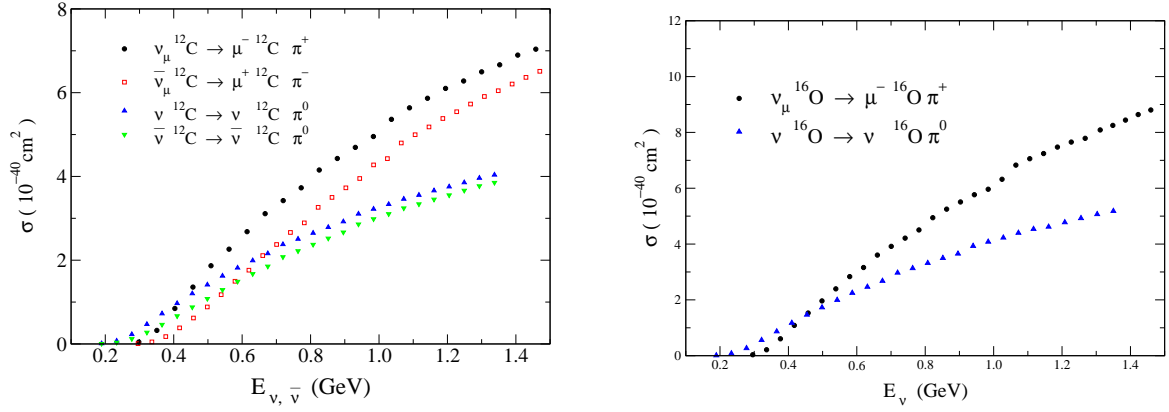

\makebox[0pt]{\includegraphics[scale=0.295]{secefcarbono.eps}\hspace{1.cm}
              \includegraphics[scale=0.295]{secefoxigeno.eps}}\\
\caption{ Muon neutrino/antineutrino CC and NC
 coherent pion production off nuclei from carbon (left) and
oxygen (right) targets as a function of the neutrino/antineutrino energy. }
\label{fig:cross}
\end{figure}
\begin{figure}[htb]
\makebox[0pt]{\includegraphics[scale=0.45]{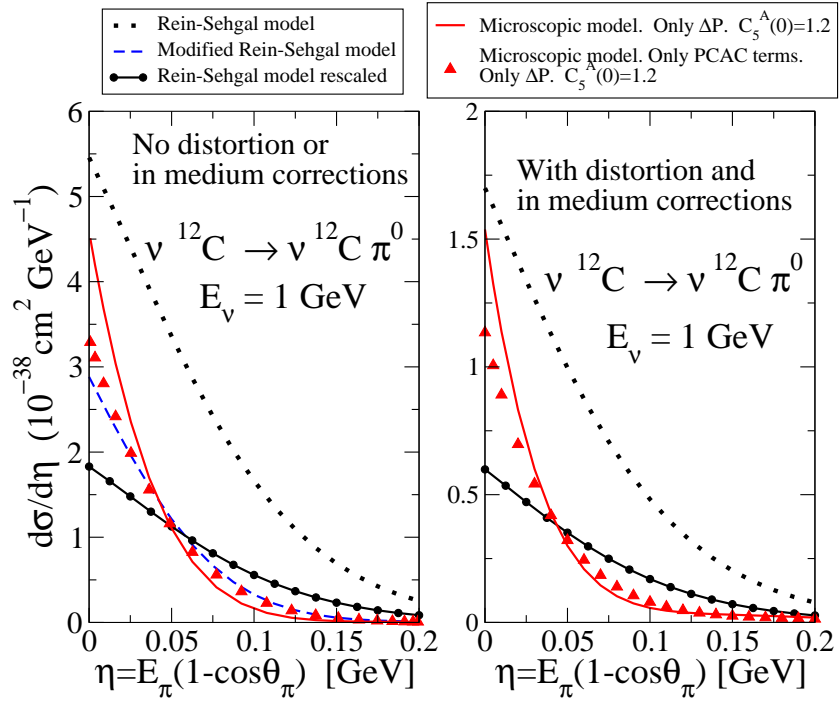}}
\caption{ Comparison of different predictions for the $\frac{d\sigma}{d\eta}$
differential cross section for NC coherent $\pi^0$ production on carbon at
$E_\nu=1\,$GeV. See text for details.}
\label{fig:cross}
\end{figure}

Let us here concentrate on the $\frac{d\sigma}{d\eta}$ differential cross
section  with  $\eta=E_\pi(1- \cos\theta_\pi)$ the variable introduced by the
MiniBooNE Collaboration in its analysis of NC coherent $\pi^0$ production~\cite{AguilarArevalo:2008xs}. In
Fig.~\ref{fig:cross} we show results for NC coherent $\pi^0$ production on
carbon at $E_\nu=1\,$GeV. In the left panel 
the  results have been evaluated without distortion or any in medium correction.
We compare the results obtained with the RS model, the one used by
experimentalists in their analyses, with the results obtained in the microscopic
model of Refs.\cite{Hernandez:2007qq,Amaro:2008hd}. In the latter case and for 
simplicity, we have only taken the dominant $\Delta P$ term and, in order to fix
the normalization, we use
$C_5^A(0)=1.2$ which is the value implicit in any PCAC-based model. Besides we show,
with triangles, results of the microscopic model eliminating all non-PCAC
contributions. Comparison of the latter with the RS model shows the
effect of the $t=0$ approximation used in that model. Cross sections are larger in the 
RS
model and what is more important this model produces a much flatter
distribution (see line ``Rein--Sehgal rescaled''). As described in Ref.~\cite{Hernandez:2009vm},
one can easily eliminate the $t=0$ approximation in the RS model by using the full non spin-flip part of 
the pion nucleon elastic cross section.
The results for this case are shown under ``Modified Rein--Sehgal'' with a
dashed line. We see the agreement with the only-PCAC-terms microscopic model is very good in this case. Another
important piece of information  can be extracted from the comparison between the full
microscopic calculation and the only-PCAC-terms one. We see the latter produces a flatter
distribution. This is mainly due to the $\frac{d\sigma}{dx\,dy\,dt\,d\phi_{k_\pi q}} 
= \frac{1}{2\pi} \frac{d\sigma}{dx\,dy\,dt}$ approximation inherent to any PCAC-based model.
In the right panel we show results that include distortion and in medium corrections. Again we
see the RS model produces larger and much flatter distributions than the microscopic
calculation due to the $t=0$ approximation and the inadequacy of the final pion distortion used. 
Besides the effect of the 
$\frac{d\sigma}{dx\,dy\,dt\,d\phi_{k_\pi q}} 
= \frac{1}{2\pi} \frac{d\sigma}{dx\,dy\,dt}$ approximation can also be seen by comparing the two
microscopic calculations.
 
\section{Conclusions}
To conclude we would like to stress the following: First,
background terms are relevant at the nucleon level and should be included in any microscopic calculation.
Even though they play a minor role in coherent production, their inclusion affects the determination of the 
$C_5^A(0)$ value which is a crucial quantity for coherent
production as it fixes the strength of the dominant $\Delta P$ contribution. Second, the 
Rein--Sehgal model 
was cleverly devised for its use at high neutrino energies and medium/large nuclei but we think
we have shown it is  inadequate for low neutrino energies and small nuclei. The use of 
the RS model in
experimental analyses could affect the determination of the rate of coherent production. We believe this is the
case for the recent determination of NC $\pi^0$ coherent production by the MiniBooNE Collaboration~\cite{AguilarArevalo:2008xs}. 
Last, some of the problems of the RS model are shared 
by all PCAC-based models: the neglect of non-PCAC terms and the implicit assumption $\frac{d\sigma}{dx\,dy\,dt\,d\phi_{k_\pi q}} 
= \frac{1}{2\pi} \frac{d\sigma}{dx\,dy\,dt}$ makes those models unable to properly determine differential cross
sections with respect to any variable that depends on the angle made by the pion and the incident neutrino. 
We have shown that for the case of the $\frac{d\sigma}{d\eta}$ differential cross section.

\vspace*{-.15cm}\begin{theacknowledgments}
  This research was supported by DGI and FEDER funds, under contracts
  FIS2008-01143/FIS, FIS2006-03438, FPA2007-65748, and the Spanish
  Consolider-Ingenio 2010 Programme CPAN (CSD2007-00042), by Junta de
  Castilla y Le\'on under contracts SA016A07 and GR12, and it is part
  of the European Community-Research Infrastructure Integrating
  Activity ``Study of Strongly Interacting Matter'' (
  HadronPhysics2, Grant Agreement n. 227431) under the Seventh
  Framework Programme of EU. M.V. wishes to acknowledge a postdoctoral fellowship form
  the Japan Society for the Promotion of Science.
\end{theacknowledgments}


\vspace{-.1cm}\begin{thebibliography}{blabla}
\bibitem{AguilarArevalo:2007it}
  A.~A.~Aguilar-Arevalo {\it et al.}  [The MiniBooNE Collaboration],
  %``A Search for electron neutrino appearance at the $\Delta m^{2} \sim
  %1$eV$^{2}$ scale,''
  Phys.\ Rev.\ Lett.\  {\bf 98}, 231801 (2007).
    %%CITATION = PRLTA,98,231801;%%

\bibitem{Hiraide:2006zq}
  K.~Hiraide  [SciBooNE Collaboration],
  %``The SciBar detector at FNAL booster neutrino experiment,''
  Nucl.\ Phys.\ Proc.\ Suppl.\  {\bf 159}, 85 (2006).
  %%CITATION = NUPHZ,159,85;%%

\bibitem{Rein:1982pf}
  D.~Rein and L.~M.~Sehgal,
  %``Coherent Pi0 Production In Neutrino Reactions,''
  Nucl.\ Phys.\  B {\bf 8223}, 29 (1983).
  %%CITATION = NUPHA,B223,29;%%

%\cite{Amaro:2008hd}
\bibitem{Amaro:2008hd}
  J.~E.~Amaro, E.~Hern\'andez, J.~Nieves and M.~Valverde,
  %``Theoretical study of neutrino-induced coherent pion production off nuclei
  %at T2K and MiniBooNE energies,''
  Phys. Rev. D {\bf 79}, 013002  (2009).
  %%CITATION = ARXIV:0811.1421;%%

%\cite{Berger:2008xs}
\bibitem{Berger:2008xs}
  C.~Berger and L.~M.~Sehgal,
  %``PCAC and coherent pion production by low energy neutrinos,''
  Phys.\ Rev.\  D {\bf 79}, 053003 (2009).
  %%CITATION = PHRVA,D79,053003;%%


%\cite{Hernandez:2009vm}
\bibitem{Hernandez:2009vm}
  E.~Hernandez, J.~Nieves and M.~J.~Vicente-Vacas,
  %``Neutrino Induced Coherent Pion Production off Nuclei and PCAC,''
  Phys.\ Rev.\ D {\bf 80},  013003 (2009).
  %%CITATION = ARXIV:0903.5285;%%

\bibitem{AguilarArevalo:2008xs}
  A.~A.~Aguilar-Arevalo {\it et al.}  [MiniBooNE Collaboration],
  %``First Observation of Coherent $\pi^0$ Production in Neutrino Nucleus
  %Interactions with $E_{\nu}<$ 2 GeV,''
  Phys.\ Lett.\  B {\bf 664}, 41 (2008).
    %%CITATION = PHLTA,B664,41;%%

\bibitem{Hernandez:2007qq}
  E.~Hern\'andez, J.~Nieves and M.~Valverde,
  %``Weak pion production off the nucleon,''
  Phys.\ Rev.\ D {\bf 76},  033005 (2007).
    %%CITATION = HEP-PH/0701149;%%



%\cite{Schreiner:1973ka}
\bibitem{Schreiner:1973ka}
  P.~A.~Schreiner and F.~Von Hippel,
  %``Nu p $\to$ mu- delta++ - comparison with theory,''
  Phys.\ Rev.\ Lett.\  {\bf 30}, 339 (1973).
  %%CITATION = PRLTA,30,339;%%

\bibitem{adler} S.L. Adler, Ann. Phys. {\bf 50} (1968) 189.


%\cite{Barish:1978pj}
\bibitem{Barish:1978pj}
  S.~J.~Barish {\it et al.},
 %``Study Of Neutrino Interactions In Hydrogen And Deuterium: Inelastic Charged
 %Current Reactions,''
  Phys.\ Rev.\  D {\bf 19} (1979) 2521.
  %%CITATION = PHRVA,D19,2521;%%

\bibitem{Radecky:1981fn}
  G.~M.~Radecky {\it et al.},
  %``Study Of Single Pion Production By Weak Charged Currents In Low-Energy
  %Neutrino D Interactions,''
  Phys.\ Rev.\  D {\bf 25} (1982) 1161.
  [Erratum-ibid.\  D {\bf 26} (1982) 3297].
  %%CITATION = PHRVA,D25,1161;%%

\bibitem{Pa04}  E.A. Paschos, J.-Y. Yu and M. Sakuda,  Phys. Rev. {\bf
  D69} (2004) 014013.
\bibitem{Alexandrou:2006mc}
  C.~Alexandrou, T.~Leontiou, J.~W.~Negele and A.~Tsapalis,
  %``The axial N to Delta transition form factors from lattice QCD,''
  Phys.\ Rev.\ Lett.\  {\bf 98}, 052003 (2007).
    %%CITATION = PRLTA,98,052003;%%

%\cite{BarquillaCano:2007yk}
\bibitem{BarquillaCano:2007yk}
  D.~Barquilla-Cano, A.~J.~Buchmann and E.~Hernandez,
  %``Axial $N\to \Delta(1232)$ and $N \to N^{\star}(1440)$ transition form
  %factors,''
  Phys.\ Rev.\  C {\bf 75}, 065203 (2007).
  [Erratum-ibid.\  C {\bf 77}, 019903 (2008)].
    %%CITATION = PHRVA,C75,065203;%%

%\cite{Graczyk:2007bc}
\bibitem{Graczyk:2007bc}
  K.~M.~Graczyk and J.~T.~Sobczyk,
  %``Form Factors in the Quark Resonance Model,''
  Phys.\ Rev.\  D {\bf 77}, 053001 (2008).
    %%CITATION = PHRVA,D77,053001;%%

\bibitem{AlvarezRuso:2007tt}
  L.~Alvarez-Ruso, L.~S.~Geng, S.~Hirenzaki and M.~J.~Vicente Vacas,
  %``Charged current neutrino induced coherent pion production,''
  Phys.\ Rev.\  C {\bf 75}, 055501 (2007).

%\cite{Oset:1987re}
\bibitem{Oset:1987re}
  E.~Oset and L.~L.~Salcedo,
  %``DELTA SELFENERGY IN NUCLEAR MATTER,''
  Nucl.\ Phys.\  A {\bf 468}, 631 (1987).
  %%CITATION = NUPHA,A468,631;%%



%\cite{Nieves:1993ev}
\bibitem{Nieves:1993ev}
  J.~Nieves, E.~Oset and C.~Garcia-Recio,
  %``A Theoretical approach to pionic atoms and the problem of anomalies,''
  Nucl.\ Phys.\  A {\bf 554}, 509 (1993).
  %%CITATION = NUPHA,A554,509;%%

%\cite{Nieves:1991ye}
\bibitem{Nieves:1991ye}
  J.~Nieves, E.~Oset and C.~Garcia-Recio,
  %``Many body approach to low-energy pion nucleus scattering,''
  Nucl.\ Phys.\  A {\bf 554}, 554 (1993).
  %%CITATION = NUPHA,A554,554;%%

\bibitem{AlvarezRuso:2007it}
  L.~Alvarez-Ruso, L.~S.~Geng and M.~J.~Vicente Vacas,
  %``Neutral current coherent pion production,''
  Phys.\ Rev.\  C {\bf 76}, 068501 (2007).
    %%CITATION = PHRVA,C76,068501;%%


%\cite{Leitner:2009ph}
\bibitem{Leitner:2009ph}
  T.~Leitner, U.~Mosel and S.~Winkelmann,
  %``Neutrino-induced coherent pion production off nuclei - revisited,''
  Phys.\ Rev.\  C {\bf 79} (2009) 057601.
  %%CITATION = PHRVA,C79,057601;%%

\end{thebibliography}
\end{document}